# Optically resonant all-dielectric diabolo nanodisks

Saddam Gafsi[1,a], Farhan Bin Tarik[1,a], Cody T. Nelson[1], and Judson D. Ryckman[1,b]

[1]Holcombe Department of Electrical and Computer Engineering, Clemson University, Clemson, South Carolina, 29634, USA

Optically resonant all-dielectric nanostructures attractively exhibit reduced losses compared to their plasmonic counterparts; however, achieving strong field enhancements at the nanoscale, especially within solid-state media, has remained a significant challenge. In this work, we demonstrate how subwavelength modifications to a conventional silicon nanodisk enable strong sub-diffractive and polarization dependent field enhancements in devices supporting anapole-like modes. We examine the electromagnetic properties of both individual and arrayed "diabolo nanodisks", which are found to exhibit $|E|^2/|E_0|^2$ enhancements in the range $\sim 10^2$-$10^4$, in the high index medium, depending on geometrical considerations. In addition to supporting a localized electric field "hot-spot" similar to those predicted in diabolo nanostructured photonic crystal cavities and waveguide designs, we identify an anti-diabolo effect leading to a broadband "cold-spot" for the orthogonal polarization. These findings offer the prospect of enhancing or manipulating light-matter interactions at the nanoscale within an all-dielectric (metal free) platform for potential applications ranging from nonlinear optics to quantum light sources, nano-sensing, nanoparticle-manipulation and active/tunable metasurfaces.

---

[a] S. Gafsi and F. Bin Tarik contributed equally to this work.

[b] jryckma@clemson.edu

A long-standing objective of nanophotonics is to enable the manipulation and control of light at the nanoscale. While plasmonic nanostructures are well suited for this task, they suffer from high Ohmic losses and limited compatibility with complementary metal oxide semiconductor (CMOS) fabrication processes. These challenges have motivated research interest in optically resonant nanostructures and metasurfaces derived from all-dielectric or high-index semiconductor platforms[1–3]. Recent advancements in this area have demonstrated methods to tailor Mie and Fano resonant phenomena[3,4], bound states in the continuum (BIC) [5,6], anapole states[7,8], Kerker effects [9,10], and more; for prospective applications ranging from wavefront engineering, to sensing and nanomanipulation, non-linear optics, and the generation of classical or quantum light. In many applications it is desirable to strongly concentrate or enhance the optical near-field in so called "hot spots". However, achieving strong sub-diffraction manipulation of light in all-dielectric media poses a significant challenge, thus motivating growing interests in dielectric-based nanostructure designs capable of tailoring electromagnetic fields at the nanoscale.

Beyond achieving resonance, one route for achieving localized field enhancements in all-dielectric media is to leverage the interfacial boundary conditions of Maxwell's equations. For example, dielectric nanoparticle dimers are known to generate near field hot spots in the nanoscaled slot-like gap between particles[11]. Similarly, slotted anapole[12] or quasi-BIC supporting structures[13] have also been proposed for strongly enhancing the electric field under the appropriate polarization. Such structures exploit continuity of the normal component of electric displacement ***D*** to generate an enhancement to the electric field ***E*** and electric field energy density, $\epsilon|E|^2$, equal to factor as large as $\epsilon_h/\epsilon_l = (n_h/n_l)^2$, where $\epsilon_h$ and $\epsilon_l$ are the relative permittivities, and $n_h$ and $n_l$ are the refractive indices of the constituent high and low index materials respectively. However, the slot effect only provides near field enhancement within the low index medium, making the hot spot inaccessible to high index solid-state media. Recently it has been shown that this limitation can be circumvented by a nanostructured design that simultaneously exploits a second boundary condition, i.e. the continuity of the *tangential* component of the electric-field [14,15]. The enforcement of this second boundary condition allows the slot-enhanced electric field to be carried inside a *high index* medium. This compound effect preserves the same $(\epsilon_h/\epsilon_l) = (n_h/n_l)^2$ field enhancement factor as the slot-effect and can enhance the electric field intensity within a high index medium by a factor as large as $(\epsilon_h/\epsilon_l)^2 = (n_h/n_l)^4$. This concept has recently been applied in the design of record low mode volume all-dielectric photonic crystal cavities[14,16–18] and low mode area all-dielectric waveguides[19], but has not yet been investigated in the context of optically resonant nanostructures and metasurfaces.

In this work, we report the design of subwavelength-engineered diabolo nanodisks and nanodisk arrays supporting strong localized electric field enhancements. By internally nanostructuring a conventional high index contrast silicon nanodisk, we create a strongly anisotropic optical response which coincides with a 'diabolo effect' enhancement to the resonant near field intensity $|\mathbf{E}|^2$, *in the high index medium*, by a factor as large as $(\epsilon_h/\epsilon_l)^2 = (n_h/n_l)^4$ at the specified design polarization. For high index contrast platforms, such as the silicon/air interface considered here, this design allows the resonant $|\mathbf{E}|^2$ to be enhanced by nearly two orders of magnitude vs. conventional optically resonant nanostructures. For the orthogonal polarization, we find that the nanostructure can give rise to an anti-diabolo effect which locally suppresses the electric field intensity to form a broadband cold spot. These findings offer the prospect of enhancing or manipulating light-matter interactions at the nanoscale within an all-dielectric (metal free) platform.

Our diabolo nanodisk design combines two distinct concepts. First, we employ multi-polar Mie resonances near the anapole condition. Anapole states emerge when the fields radiated by the co-located electric and toroidal dipoles cancel each other in the far-field through destructive interference in a special type of Fano resonance[7,8,20]. It has been reported that the field enhancement is maximized not necessarily at the anapole frequency, but rather in its vicinity at the frequency corresponding to the higher quality factor mode contributing to the Fano resonance [20]. Strong field confinement in an "anapole-like" mode distribution may still be obtained even for structures with imperfect far-field cancellation or a non-zero minimum scattering cross-section[21]. Secondly, the internal diabolo nanostructure of our nanodisk is aimed at exploiting the vectorial nature of light to produce an electric field distribution that is constrained by satisfying both the continuity of normal electric displacement ***D*** and tangential electric field ***E*** at the dielectric boundaries, thus producing a strong field enhancement at the center of the disk which is carried into the high index medium. We note that the term "bowtie" is sometimes used in the literature for devices leveraging this concept[14,16–18], however we prefer the term "diabolo" since bowtie structures have historically featured nanoscale gaps between tips[22], whereas diabolo structures are historically connected at the tips[23–25]. To validate our nanostructure design concept, we theoretically examined the electromagnetic properties and near-field enhancement of isolated and arrayed diabolo nanodisks via 3D finite difference time domain (FDTD) modelling. We then experimentally prototyped diabolo nanodisk arrays in a 220 nm device layer silicon-on-insulator (SOI) platform, with an air top cladding and 2 μm buried oxide cladding and characterized their anisotropic resonant response.

Figure 1 shows an illustration of the diabolo nanodisk and its predicted electromagnetic behavior in vacuum. As shown in Fig. 1(a), a diabolo nanodisk is comprised of a conventional circular nanodisk[7,26] with radius $R$ that is modified by etching two diagonally oriented square boxes of width $w$ and separation $h$. Fig. 1(b) shows an SEM image of an example diablo nanodisk particle after fabrication with electron-beam lithography and dry etching in a 220 nm thick crystalline silicon (100) device layer. This structure is realized with dimensions $R$ = 230 nm, $w$ = 110 nm, and $h$ = 25 nm. From SEM we also extract the radius of curvature at the corners of the etched boxes to be ~18 nm. It is important to account for physically realistic radius of curvature dimensions when modelling electromagnetic fields near corners, otherwise singularities may arise which yield non-convergent and/or non-physical results[15,19,27].

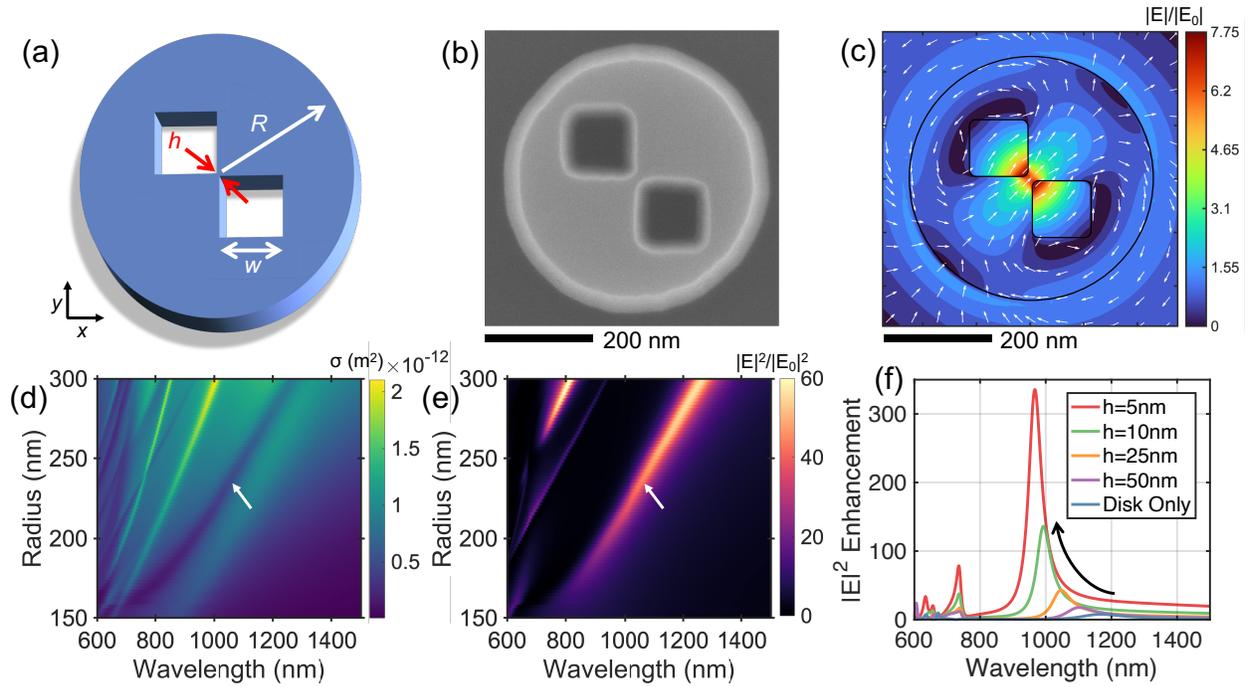

**Figure 1.** (a) Illustration and (b) SEM of an all-dielectric diabolo nanodisk. (c) Simulated E-field profile for the optically resonant anapole-like mode of a single diabolo nanodisk in vacuum with dimensions corresponding to (b). (d) Simulated scattering cross-section and (e) electric field intensity enhancement vs. nanodisk radius R, where w = 110 nm, h = 25 nm, and silicon thickness = 220 nm. The white arrows correspond to the device simulated in part (c). (f) Predicted $|E|^2$ enhancement of the isolated diabolo nanodisk (w = 110 nm, R = 230 nm) as a function of $h$ considered alongside a reference solid nanodisk.

circular nanodisk[7,26] with radius $R$ that is modified by etching two diagonally oriented square boxes of width $w$ and separation $h$. Fig. 1(b) shows an SEM image of an example diablo nanodisk particle after fabrication with electron-beam lithography and dry etching in a 220 nm thick crystalline silicon (100) device layer. This structure is realized with dimensions $R$ = 230 nm, $w$ = 110 nm, and $h$ = 25 nm. From SEM we also extract the radius of curvature at the corners of the etched boxes to be ~18 nm. It is important to account for physically realistic radius of curvature dimensions when modelling electromagnetic fields near corners, otherwise singularities may arise which yield non-convergent and/or non-physical results[15,19,27].

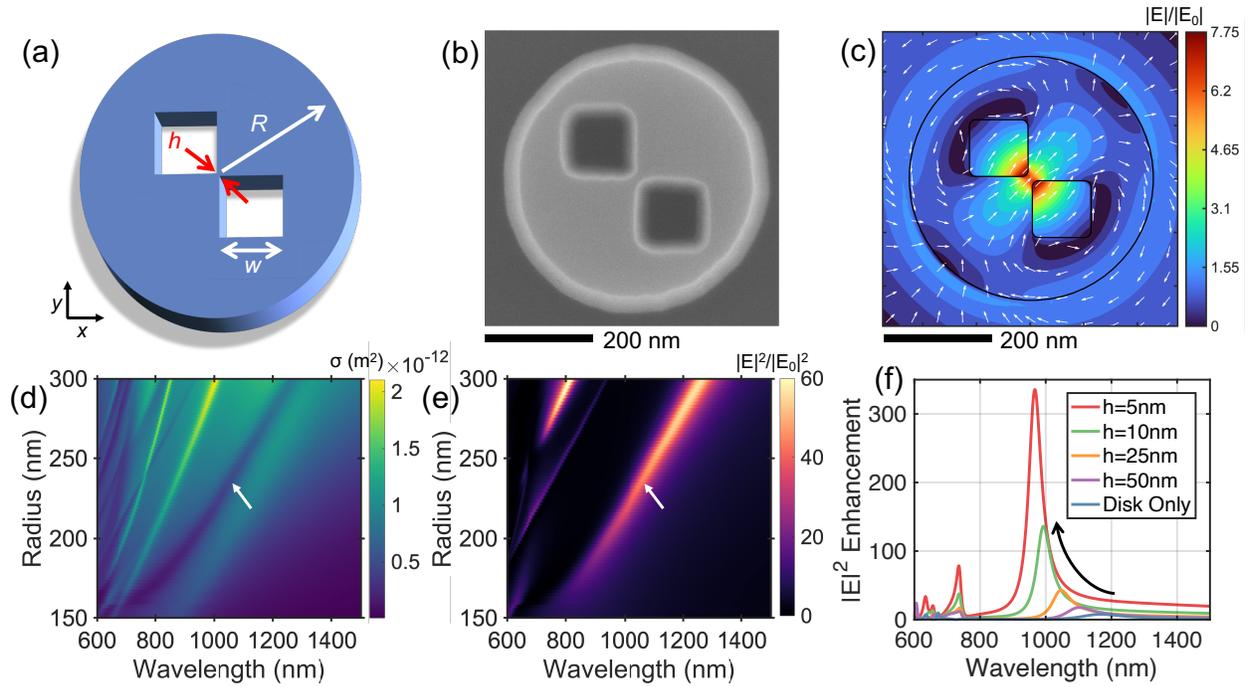

**Figure 1.** (a) Illustration and (b) SEM of an all-dielectric diabolo nanodisk. (c) Simulated E-field profile for the optically resonant anapole-like mode of a single diabolo nanodisk in vacuum with dimensions corresponding to (b). (d) Simulated scattering cross-section and (e) electric field intensity enhancement vs. nanodisk radius R, where w = 110 nm, h = 25 nm, and silicon thickness = 220 nm. The white arrows correspond to the device simulated in part (c). (f) Predicted $|E|^2$ enhancement of the isolated diabolo nanodisk (w = 110 nm, R = 230 nm) as a function of $h$ considered alongside a reference solid nanodisk.

To demonstrate the electromagnetic properties of the diabolo nanodisk, we first performed 3D FDTD simulation of single particles in vacuum under normal incidence plane wave illumination from a source linearly polarized at +45 degrees, corresponding to an electric field oriented along the $(\hat{x} + \hat{y})$ direction. The minimum FDTD mesh size in each direction are given by $(dx, dy, dz)$ = (2, 2, 5) nm. Excitation of a diabolo nanodisk with the same dimensions as the structure in Fig. 1(b) reveals a resonant mode with an anapole-like electric field distribution near 1050 nm featuring a strong field enhancement at the center of the disk as depicted in Fig. 1(c). The localized field enhancement is achieved not solely in a low index medium, as in the case of nanodisk dimers or slotted/gapped structures, but also in the high index medium at the center of the disk.

In Figs. 1(d, e) we present the simulated scattering cross-section and electric field intensity enhancement for individual diabolo nanodisks with $h$ = 25 nm and $w$ = 110 nm for varying $R$ values. Similar to nanodisks supporting conventional anapole modes[7,28], we observe a characteristic minimum in the scattering cross-section alongside a resonant field enhancement which redshifts with increasing radii. For $R \geq 225$ nm a single anapole-like mode is indicated to resonate in the near infrared at wavelengths $\lambda_0 \geq 1000$ nm; whereas at larger radii, e.g. $R \geq 275$ nm, the structure is found to support a second anapole mode exhibiting field enhancement at shorter wavelengths, e.g. $\lambda_0 \geq 750$ nm. In this study, we focus primarily on the properties of the fundamental anapole-like mode of our diabolo nanodisks; however, we note that our key findings and design principles are expected to similarly apply toward higher order anapole modes.

In Figs. 1(f) we compare the simulated $|E|^2$ enhancement for individual diabolo nanodisks with $R$ = 230 nm and $w$ = 110 nm for varying $h$ values. These results are presented alongside the corresponding properties for a solid reference nanodisk ($R$ = 230 nm) with no internal nanostructuring. Initially the solid nanodisk exhibits a minimum in scattering near 1130 nm and a maximum resonant $|E|^2/|E_0|^2$ enhancement, as measured in silicon at the center of the nanodisk, of ~8x near $\lambda_0 = 1170$ nm. Introduction of the etched diabolo nanostructure within the nanodisk lowers the average refractive index of the particle and blueshifts the resonance wavelength. As $h$ decreases from 50 nm toward 5 nm the resonance blueshifts further and significant enhancement in the maximum resonant $|E|^2$ is predicted, rising from $|E_{max}|^2/|E_0|^2 = 18$ for the case $h = 50$ nm ($\lambda_0 = 1103$ nm), to $|E_{max}|^2/|E_0|^2 = 335$ for the case $h = 5$ nm ($\lambda_0 = 967$ nm). These results demonstrate that the magnitude of the diabolo effect is strongly dependent on $h$ and indicates the $|E|^2$ enhancement approaches the theoretical maximum enhancement factor $(n_h/n_l)^4$ as $h$ decreases, an effect which is similarly predicted in low mode area diabolo waveguides[19] and low mode volume photonic crystal cavities[18,29].

The extreme near-field intensity and highly localized optical interaction with silicon is further evidenced by the large sensitivity of the resonant wavelength with respect to nanoscale variations in $h$, where $d\lambda_0/dh \approx 5$ nm/nm is observed for $h < 25$ nm. This suggests that diabolo nanodisk geometry could prove effective for applications such as nanosensors or dynamic metasurfaces where high sensitivity to perturbations within nanoscaled volumes is required. The large $|E|^2/|E_0|^2$ is also desirable for modifying the spontaneous emission lifetime of quantum emitters[30], for enhancing non-linear effects[31], and for optical trapping of nanoscaled objects[6]. In many such applications it is preferred to work with arrayed nanostructures or metasurfaces rather than isolated/single nanostructures, as considered in the following section.

Periodic metasurfaces provide a convenient interface between free-space and on-chip optics. Moreover, coherent interactions within a uniform periodic lattice can further enhance the $|E|^2$ owing to the formation of Fano or guided mode resonances [4,9]. In Figure 2 we present the predicted characteristics of diabolo nanodisk arrays arranged into a square lattice with period $P = 675$ nm.

Here we consider structures of varying *h*, from 2 nm to 100 nm, in a 220 nm silicon device layer with *R* = 230 nm, an air top cladding, and a 2 μm thick buried oxide (BOX) bottom cladding residing on the silicon handle layer. As shown in Fig. 2(a) we observe a pronounced resonance in the reflectivity spectra, which coincides with a strong $|E|^2/|E_0|^2$ enhancement as measured at the center of the disk and shown in Fig. 2(b). Whereas an isolated particle with nanoscaled $h \approx 10$ nm exhibits an electric field intensity enhancement on the order of ~$10^2$ (or ~20 dB), the periodic metasurface supports localized enhancements in the range ~$10^3$-$10^4$ (or ~30 to 40 dB) as indicated in Fig. 2(c).

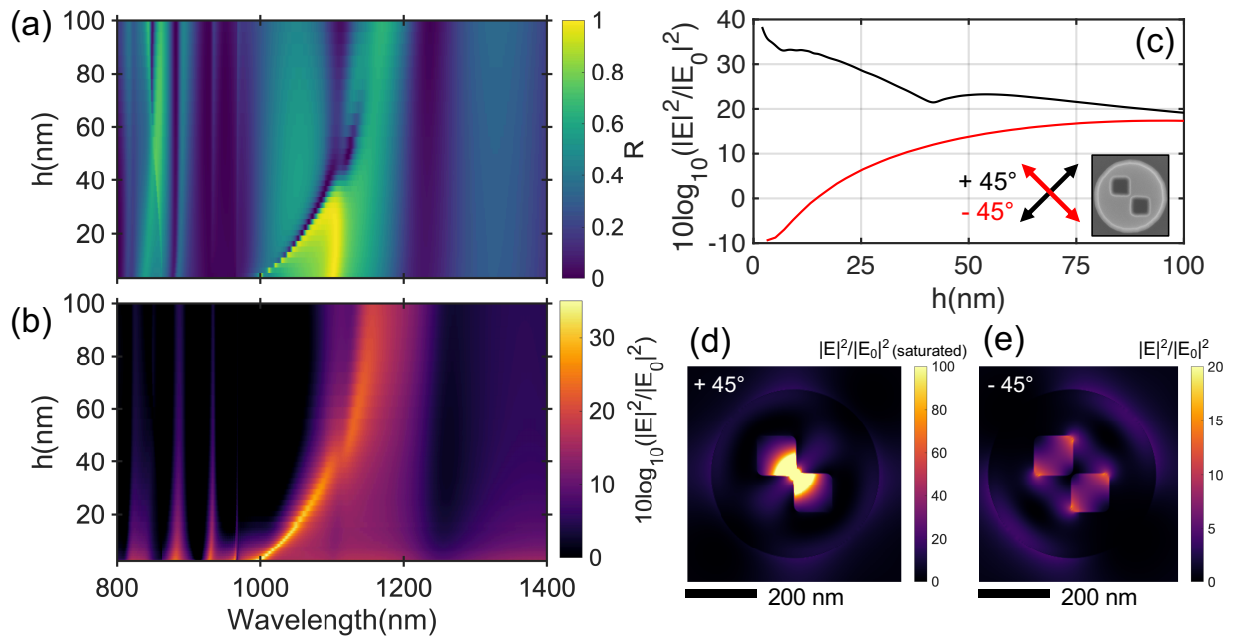

**Figure 2.** Simulated (a) reflectance and (b) $|E|^2$ enhancement (log scale) of periodic diabolo nanodisk metasurface in 220 nm SOI with period *P* = 675 nm, *R* = 230 nm, *w* = 110 nm shown as a function of *h*, in the case of +45 deg polarization. (c) Extracted polarization dependent $|E|^2$ enhancement (log scale) at the center of the nanodisk, as measured at the resonance wavelength for +45 deg polarization. (d) Resonant $|E|^2$ profile for the case h = 5 nm (saturated for enhancements >100 to show detail) for +45 deg polarization, and (e) the corresponding $|E|^2$ profile for -45 deg polarization.

We further evaluated the polarization dependance of the diabolo nanodisk's near field enhancement by evaluating $|E|^2/|E_0|^2$ at the center of the disk for the orthogonal polarization

(-45 degrees). As shown in Fig. 2(c), the electric field intensity is increasingly suppressed with decreasing $h$. This localized near-field suppression at the center of the disk is attributed to an anti-diabolo effect. Whereas the aforementioned continuity of normal displacement and tangential electric field in the diabolo structure gives rise to a hot spot for the +45 degree polarization, it effectively gives rise to a *cold spot* for the -45 degree polarization. This effect is shown in the near field $|\mathbf{E}|^2$ distributions depicted in Fig. 2(d) and (e) respectively, which illustrate the case where $h = 5$ nm. In this scenario, the contrast in electric field intensity between orthogonal polarizations, as measured at the center of the nanodisk, is greater than four orders of magnitude. This represents a significant departure compared to the case of a conventional nanodisk, which has radial symmetry and zero polarization contrast; or a diabolo nanodisk with large $h > 75$ nm, where the contrast is less than one order of magnitude. This strongly anisotropic near field response opens the possibility of using polarization control to strongly modulate light-matter interactions at the nanoscale.

Fig. 3(a) shows an SEM image of an experimentally realized diabolo nanodisk array that was fabricated by electron beam lithography and reactive ion etching in a 220 nm SOI platform (Applied NanoTools Inc.). The nanostructure dimensions were estimated by SEM as: $R = 230$ nm, $w = 110$ nm, $h = 25$ nm, with a period $P = 675$ nm. The BOX layer thickness was estimated to be ~1.94 μm from fitting thin-film reflectance spectra. To characterize our diabolo nanodisk array and verify its anisotropic resonant response, we performed polarization resolved reflectance measurements in an optical microscope (NA = 0.4) using visible and near-infrared spectrometers (OceanOptics) with a resolution bandwidth of ~10 nm.

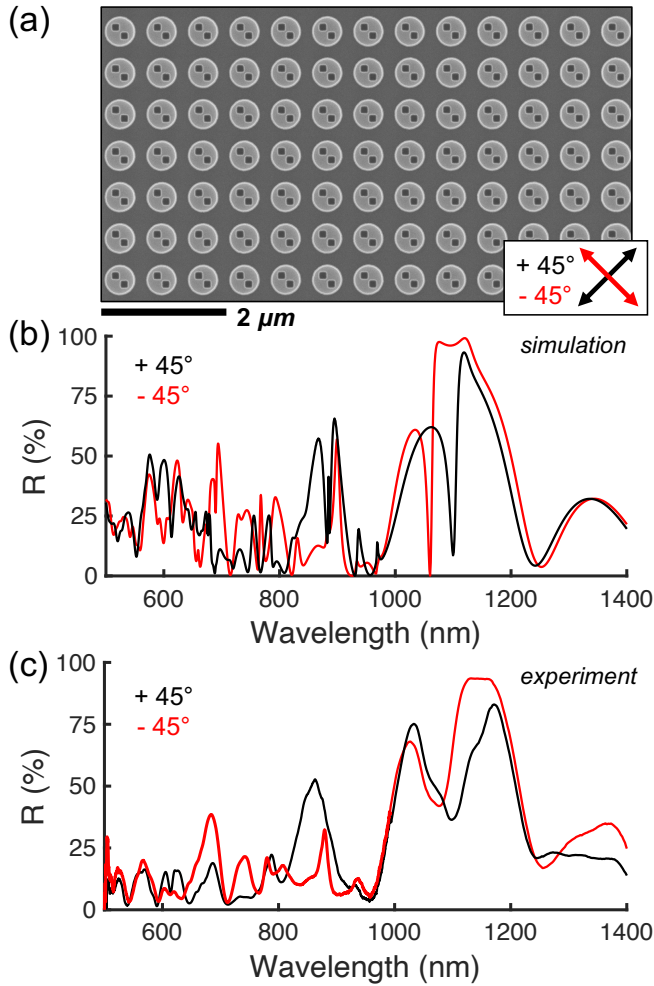

**Figure 3.** (a) SEM image of an experimentally realized diabolo nanodisk array. (b) Simulated reflectance for both linear polarizations revealing an anisotropic anapole-like mode resonance near ~1100 nm. (c) Experimentally measured reflectance as measured in an optical microscope (NA=0.4), confirming the anisotropic resonant response.

As shown in Fig. 3(b), simulations based on our experimental geometry predict the presence of a resonant anapole-like mode near ~1100 nm for +45 degree polarization, while the -45 degree polarization supports a broad reflectivity peak in this region alongside a blue shifted resonance indicative of reduced field confinement in silicon. Each resonance exhibits an asymmetric Fano line-shape with a quality factor on the order of ~75. The experimental spectra obtained for each polarization are displayed in Fig. 3(c) and exhibit good overall agreement with the device simulation. The measurements confirm the diabolo nanodisk supports a strongly anisotropic resonant response. We note that the measured resonance Q factors are ~3x lower than simulated values and appear to be limited by two main factors: (1) interrogation of the array in the microscope averaging across a cone of angles defined within the collection NA[32], and (2) the ~10 nm resolution bandwidth of the spectrometer.  We note that in general the Q factor will be limited by wavelength dependent material absorption properties and modal confinement; whereas in windows of transparency, Q factor improvements could be expected from design optimization[33].

In this work we've introduced the diabolo nanodisk design concept as a means for locally enhancing the electric field in an all-dielectric system with high index contrast.  Unlike slotted nanodisk structures[12], where field enhancement is provided in the low index medium only, the diabolo design makes strong field enhancements accessible to the high refractive index medium. In future applications and experiments it would be attractive to embed solid-state quantum emitters within the dielectric medium, to leverage non-linearities derived from the nanodisk medium or hybrid materials which may be co-integrated within the diabolo region, or use the hot spot and strong optical gradient generated at the center of the disk manipulate or trap nanoscaled objects. Meanwhile, the strong anisotropic response of our design offers the opportunity to tailor light-matter interactions at the nanoscale by controlling the input polarization state.  Given that the diabolo design concept provides an $|E|^2$ enhancement that scales with $(n_h/n_l)^4$, it would further be attractive to consider alternative material platforms and spectral regions supporting ultra-high refractive index contrasts[21].

In conclusion, we reported how simple sub-wavelength scale modifications to a well-known structure, e.g. the dielectric nanodisk, can be utilized to manipulate electromagnetic fields at the nanoscale.  Our findings demonstrate that the diabolo design concept, previously considered in the contexts of photonic crystal cavity and waveguide design, can equally be applied to optically resonant nanostructures and metasurfaces.  This opens an important design tool to this class of devices and offers potential to enhance light-matter interactions for applications ranging from nonlinear optics, to quantum light sources, nano-sensing, nanoparticle-manipulation and more.

**ACKNOWLEDGMENTS**

This work was supported by the Air Force Office of Scientific Research (AFOSR) under Grant No. FA9550-19-1-0057 (G. Pomrenke).